\documentclass[
twocolumn,groupedaddress,floatfix,preprintnumbers
amsfonts,amssymb,amsmath,
a4paper,twocolumn,brazil,english]{revtex4-1}
\usepackage[T1]{fontenc}
\usepackage[latin1]{inputenc}
\usepackage{verbatim}
\usepackage{bm}
\usepackage{color}
\usepackage{amsthm}
\usepackage{stmaryrd}
\usepackage{esint}
\usepackage{lineno}
\usepackage{graphicx}

\pdfpageheight\paperheight
\pdfpagewidth\paperwidth

\usepackage{babel}

\begin{document}
\selectlanguage{english}

\title{Effect of hybridization symmetry on topological phases of odd-parity multiband superconductors}
\author{T.O. Puel$^{1}$}
\email{tharnier@me.com}
\author{P.D. Sacramento$^{1,2}$}
\author{M. A. Continentino$^{1}$}
\affiliation{$^{1}$Centro Brasileiro de Pesquisas F\' isicas­, Rua Dr. Xavier Sigaud,
150, Urca 22290-180, Rio de Janeiro, RJ, Brazil }
\affiliation{$^{2}$CeFEMA, Instituto Superior T\' ecnico, Universidade de Lisboa,
Av. Rovisco Pais, 1049-001 Lisboa, Portugal}
\date{\today}

\begin{abstract}
We study two-band one-dimensional superconducting chains of spinless
fermions with inter and intra-band pairing. These bands hybridize
and, depending on the relative angular momentum of their orbitals, the
hybridization can be symmetric or anti-symmetric. The self-consistent competition between intra
and inter-band superconductivity and how it is affected by the symmetry
of the hybridization is investigated. 
In the case of anti-symmetric hybridization the intra and inter-band pairings
do not coexist while in the symmetric case they do coexist and
the interband pairing is shown
to be dominant. 
The topological properties of the model are obtained through the topological
invariant winding number and the presence of edge states.
We find the existence of a topological phase
due to the inter-band superconductivity and induced by symmetric hybridization.
In this case we find a characteristic $4\pi$-periodic Josephson current.
In the case of anti-symmetric hybridization we also find a $4\pi$-periodic Josephson current in the gapless inter-band superconducting phase,
recently identified to be of Weyl-type.
\end{abstract}

\maketitle

\section{Introduction}

Multiband models for the superconducting state and their topological
properties have received increasing attention recently \cite{Kortus-2007,Xi-2008,Yang-Wang-Xiang-Li-Wang-2013,Yin-Baarsma-Heikkinen-Martikainen-Torma-2015,Nourafkan-Kotliar-Tremblay-2015,Watanabe-Yoshida-Yanase-2015}.
This consideration has been important to explain many important effects
in topological systems. For instance, topological semimetals \cite{Sun-Liu-Hemmerich-Sarma-2012}
and chiral superfluidity \cite{Liu-Li-Wu-Liu-2014} have been predicted
in multiorbital models where orbitals with different symmetries interact.
Two component fermionic systems with occupied $s$ and $p$ orbital
states were shown to have a rich phase diagram in both one and two
dimensions \cite{Yin-Baarsma-Heikkinen-Martikainen-Torma-2015}.
A general connection between multiband and multicomponent superconductivity
has also been made \cite{Tanaka-2015}. Topological properties in
three-band models were also studied \cite{Stanev-2015,Go-Park-Han-2013,Lee-Park-Go-Han-2013,He-Moore-Varma-2012}.

It is well known that the Kitaev model \cite{Kitaev-2001,Kitaev-2003,Alicea-2012}
-- anti-symmetric pairs of spinless fermions in 1D -- is the simplest
model that exhibits a topological phase with Majorana modes in the
ends of a $p$-wave chain,
depending on the state of the system. The topological non-trivial 
phase presents Majorana fermions at its ends. Otherwise, the chain
is in a superconducting phase with trivial
topological properties and has no end states\cite{Alicea-2012}. 
An extension of this effective spinless fermions model for a multiband hybridized system
comprised of the Su, Schrieffer and Heeger (SSH) model\cite{Su-Schrieffer-Heeger-1979}
and the Kitaev model was done in Ref. [\onlinecite{Wakatsuki-Ezawa-Tanaka-Nagaosa-2014}],
where topological properties are discussed showing edge states that are
of Majorana and fermionic types.

Triplet superconductivity is rare in nature. Thus, the pursuit of
alternatives to create triplet superconductivity lead to engineering
a topological insulating chain (made with strong spin-orbit material)
in proximity of a normal superconductor and in the presence of an applied magnetic field
\cite{Sato-Takahashi-Fujimoto-2009,Sau-Lutchyn-Tewari-Sarma-2010}.
On the other hand, triplet pairing has been found to be physically
realizable in some systems. 
In Ref. [\onlinecite{Watanabe-Yoshida-Yanase-2015}]
it was shown that odd-parity superconductivity occurs in superconducting
(SC) multilayers, where this state is a symmetry-protected topological
state. In addition, triplet pairing is found in $^{3}$He \cite{Vollhardt-Wolfle-2013}
and in Sr$_{2}$RuO$_{4}$\cite{Mackenzie-Maeno-2003}, as well as
in some rare noncentrosymmetric systems \cite{Sato-Fujimoto-2009}.
Triplet pairing was also studied in the context of extended Hubbard chain \cite{Sun-Chiu-Hung-Wu-2014}. 

Motivated by the recently discussed topological characters of multiband
models\cite{Yin-Baarsma-Heikkinen-Martikainen-Torma-2015,Watanabe-Yoshida-Yanase-2015},
and based on the simplest model that describes the topological properties
of a chain of spinless fermions, we study the Kitaev model with two
orbital-bands. We include and discuss inter- and intra-band superconducting
couplings. A characteristic feature of multiband systems is the hybridization
between the different orbitals. This arises from the superposition
of the wave functions of these orbitals in different sites.
It can have distinct symmetry properties depending on the orbitals
involved. If this mixing involves orbitals with angular momenta that
differ by an odd number, hybridization turns out to be anti-symmetric,
i.e., in real space we have $V_{ij}=-V_{ji}$ or in momentum, $k$-space,
$V(-k)=-V(k)$. Otherwise hybridization is symmetric respecting inversion
symmetry in different sites \cite{Deus-Continentino-Caldas-2015}.

The bulk-edge correspondance guarantees that in the topological phases
there are subgap edge states. In the case of a topological superconductor,
zero energy Majorana modes are predicted to appear and great effort
has been devoted to prove their existence. Methods that provide signatures
of their presence have been proposed and experimentally tested via for instance
tunneling experiments \cite{Law-Lee-Ng-2009,Wimmer-Akhmerov-Dahlhaus-Beenakker-2011}, interferometry
\cite{Alicea-2012},
point contacts using the Andreev
reflection \cite{Burmistrova-Devyatov-Golubov-Yada-Tanaka-2013}
through the detection of zero-bias peaks
\cite{Das-Ronen-Most-Oreg-Heiblum-Shtrikman-2012},
using the quantum waveguide theory \cite{Araujo-Sacramento-2009}
which gives the
correct bulk-edge correspondence \cite{Silva-Araujo-Sacramento-2015}
and fractional Josephson currents \cite{Kitaev-2001,Alicea-2012,Sato-Fujimoto-2016}.
Also signatures of the Majorana states may be found in bulk measurements such as
the imaginary part of frequency dependent Hall conductance \cite{Ojanen-Kitagawa-2013}
and the d.c. Hall conductivity itself \cite{Sacramento-Araujo-Castro-2014}.

The existence of topological phases is detected in this work numerically calculating the winding
number and by showing the existence of edge states at the ends of the chain.
In addition, we calculate the Josephson current accross the junction between two
superconductors to identify regimes where the periodicity of the Josephson current
on the phase differences between the superconductors (original proposal by Kitaev\cite{Kitaev-2001}) or the equivalent situation of a superconducting ring 
 threaded by a magnetic flux and interrupted by an insulator changes from the usual value of $2 \pi$ to a $4 \pi$ value \cite{Lutchyn-Sau-Sarma-2010}.
As shown before 
\cite{Kitaev-2001,Lutchyn-Sau-Sarma-2010,Xu-Fu-2010,Jiang-Pekker-Alicea-Refael-Oreg-Oppen-2011,Kwon-Sengupta-Yakovenko-2004,Kwon-Yakovenko-Sengupta-2004,Lutchyn-Sau-Sarma-2010,Oreg-Refael-Oppen-2010,Cheng-Lutchyn-2012,Dominguez-Hassler-Platero-2012,Law-Lee-2011,Gibertini-Taddei-Polini-Fazio-2012,Beenakker-Pikulin-Hyart-Schomerus-Dahlhaus-2013}
the existence of the Majoranas at the edges allows tunneling of a
single fermion at zero-bias leading to a $4\pi-$periodic current in contrast to the usual
Cooper pair transport accross the junction which leads to the usual $2\pi-$periodic
current.
Experimental realization to detect $4\pi$-periodic Josephson junction has been presented in Ref. 
[\onlinecite{Lee-Jiang-Houzet-Aguado-Lieber-Franceschi-2014}]
and an application to multiband systems has recently been presented in Ref.
[\onlinecite{Alase-Cobanera-Ortiz-Viola-2016}].

This paper is organized as follows. In section \ref{sec:the model}
we define the general Hamiltonian including symmetric and anti-symmetric
hybridization. Also we proceed with the self-consistent calculations of the superconducting
order parameters related to the competition between the 
intra- and inter-band pairings. The topological
properties of the model are discussed in section \ref{sec:Topology}.
We show a general calculation of the winding number when particle-hole
symmetry is present in a $4\times4$ Bogoliubov-de Gennes (BdG) Hamiltonian.
Also, we calculate the energy spectrum
of a finite one-dimensional chain. The differences between trivial
and topological phases are discussed from the perspective of zero-energy
states. 
We also make the equivalence of the
topological regimes with the $4\pi$ periodicity of the Josephson current.
Finally, in section \ref{sec:Conclusions} we present the
conclusions and review the main results.

\begin{figure*}[th]
\includegraphics[width=0.32\textwidth]{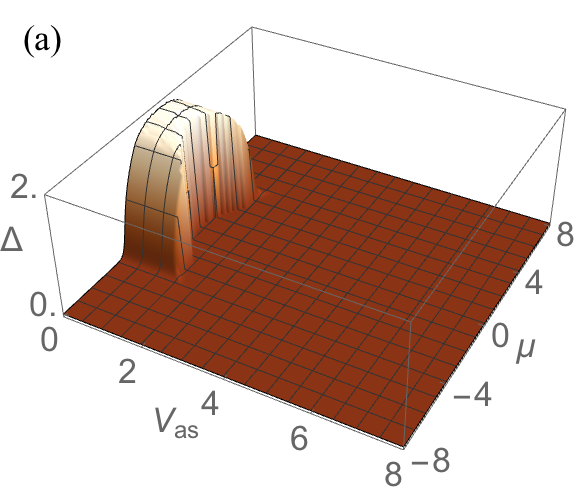}
\includegraphics[width=0.32\textwidth]{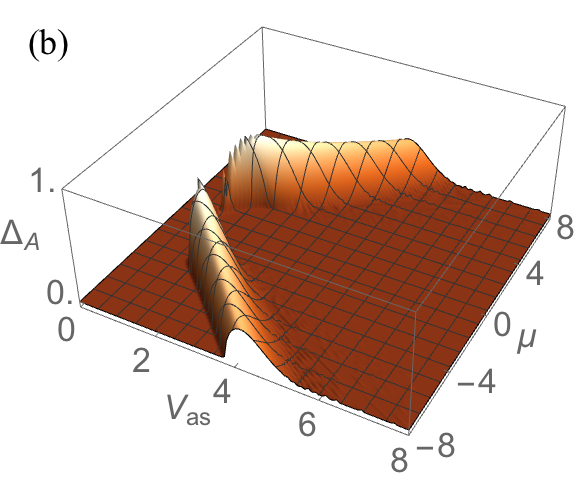}
\includegraphics[width=0.32\textwidth]{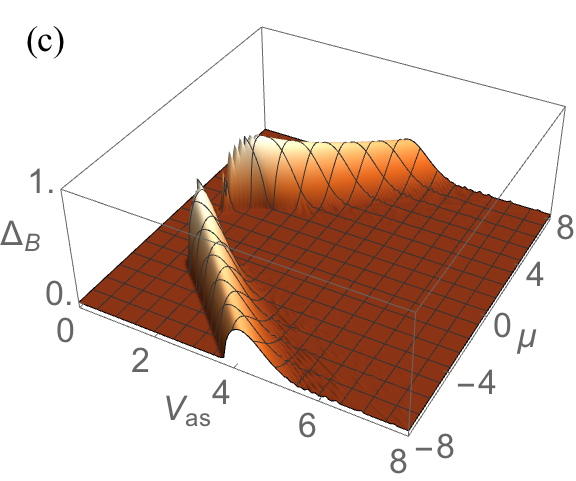}
\includegraphics[width=0.32\textwidth]{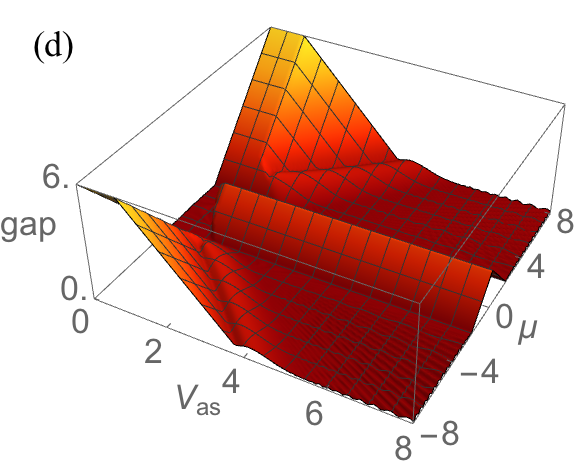}
\includegraphics[width=0.30\textwidth]{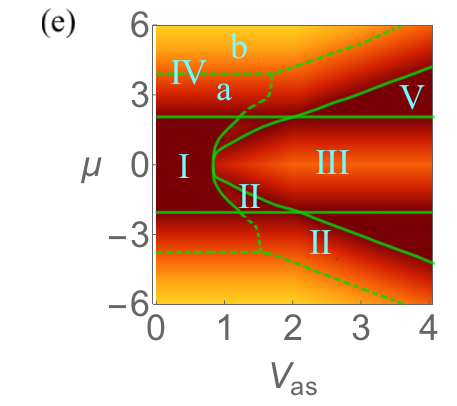}
\caption{The first row shows the self-consistent solutions of the
superconducting parameters, considering
anti-symmetric hybridization. The order parameters calculated are
the inter-band ($\Delta$) and the intra-band ($\Delta_{A,B}$) ones.
For instance, according to the anti-symmetric hybridization, $A$
and $B$ could be the orbitals $s$ and $p$. For these results we
set $g/2=g_{\text{A}}=g_{\text{B}}=1.7$.
Second row shows the corresponding energy spectrum gap and the phase diagram.
Phase I is a gapless inter-band superconducting phase.
II is a gapped intra-band superconducting phase.
III is a topological insulating phase. 
IVa shows a trivial gapped inter-band SC. 
IVb is a trivial insulating phase.
Finally, V is a metallic
phase. The phase diagram is symmetric around $\mu=0$.}
\label{self-consist anti-sym hybrid}
\end{figure*}

\begin{figure*}[th]
\includegraphics[width=0.32\textwidth]{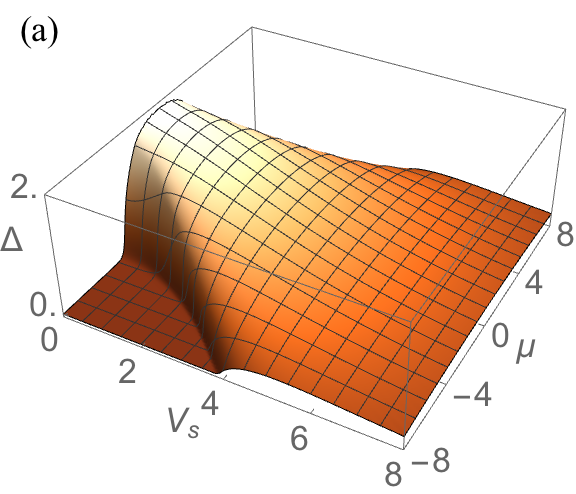}
\includegraphics[width=0.32\textwidth]{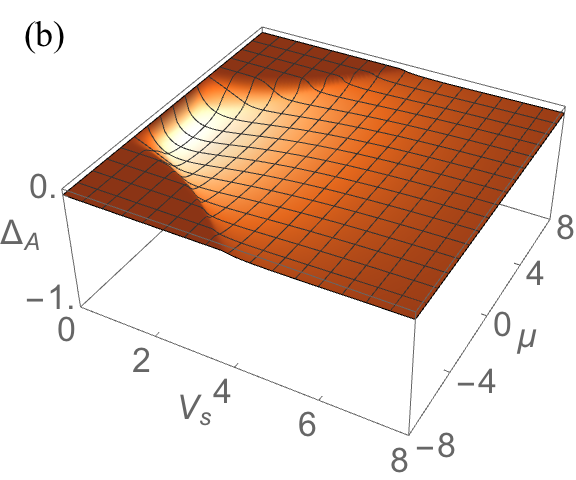}
\includegraphics[width=0.32\textwidth]{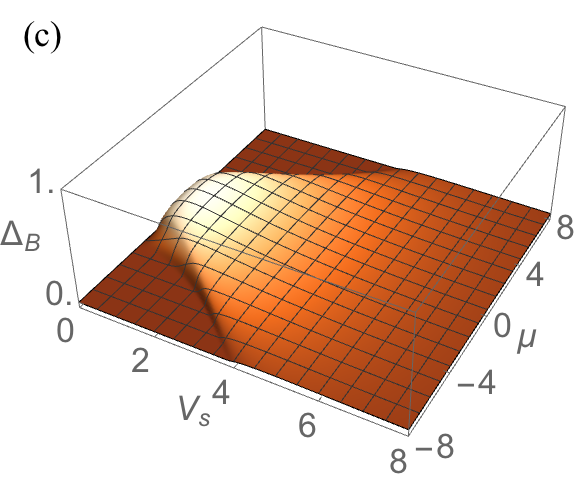}
\includegraphics[width=0.32\textwidth]{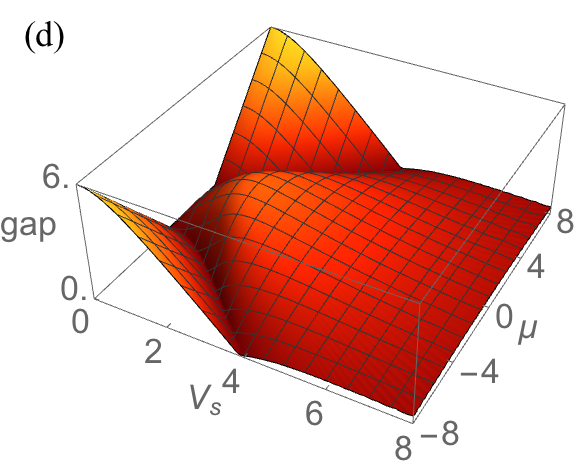}
\includegraphics[width=0.30\textwidth]{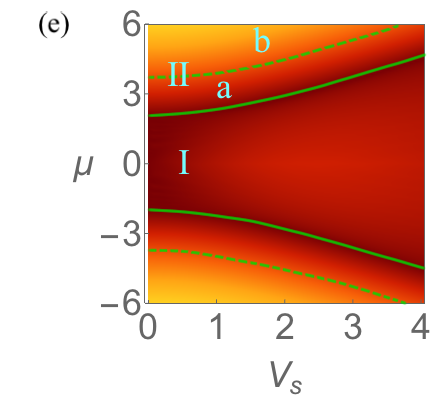}
\caption{First row shows the self-consistent solutions of the superconducting parameters, considering
symmetric hybridization. The order parameters calculated are the inter-band
($\Delta$) and the intra-band ($\Delta_{A,B}$). For instance,
according to the symmetric hybridization, $A$ and $B$ could be the
orbitals $s$ and $d$. For these results we set $g/2=g_{\text{A}}=g_{\text{B}}=1.7$.
Second row shows the corresponding spectral gap and the phase diagram.
Phase I carries both types of pairings and
has non-trivial topological properties. Phase IIa is a gapped superconducting
phase also with both inter- and intra-band pairings, but trivial topological
properties. Phase IIb is a normal insulator. The phase diagram is
symmetric around $\mu=0$.}
\label{self-consist sym hybrid}
\end{figure*}

\section{Model and self-consistent calculations \label{sec:the model}}

We consider a two-band superconductor with hybridization and triplet
pairing in 1D, i.e., a chain of sites supporting two orbitals, let's
say orbitals $A$ and $B$. The pairing between fermions may exist
on different bands (inter-band) or in each band (intra-band) and are
always of $p$-wave type, in the sense that pairs of spinless fermions
are spatially anti-symmetric. The problem can be viewed as a generalization
of the Kitaev model to two orbitals. We also have the hybridization
term between the orbitals A and B that may be symmetric or anti-symmetric.
The simplest Hamiltonian in momentum space that describes those types
of superconductivity and hybridization may be written as $\mathcal{H}=\mathcal{H}_{0}+\mathcal{H}_{h}+\mathcal{H}_{SC}$
where the kinetic part is 
\begin{equation}
\mathcal{H}_{0}=\sum_{k}\left\{ \left(\varepsilon_{k}^{A}-\mu\right)a_{k}^{\dagger}a_{k}+\left(\varepsilon_{k}^{B}-\mu\right)b_{k}^{\dagger}b_{k}\right\} ,
\end{equation}
where $a_{k}^{\dagger}\left(b_{k}^{\dagger}\right)$ is the creation
operator of spinless fermion at $A\left(B\right)$-band with momentum
$k$. Also, $\mu$ is the chemical potential and we choose $\varepsilon_{k}^{A}=-\varepsilon_{k}^{B}=2t\cos\left(k\right)\equiv\varepsilon_{k}$
where $t$ is the hopping amplitude. The hybridization term is
\begin{eqnarray}
\mathcal{H}_{h} & = & \sum_{k}\left\{ V\left(k\right)a_{k}^{\dagger}b_{k}-V\left(-k\right)b_{-k}a_{-k}^{\dagger}+\text{h.c.}\right\} ,
\end{eqnarray}
where $V\left(k\right)=2i V_{\text{as}} \sin\left(k\right)\equiv V_{\text{as},k}$
if the hybridization is anti-symmetric or $V\left(k\right)=2 V_{\text{s}} \cos\left(k\right)\equiv V_{\text{s},k}$
if the hybridization is symmetric, and $V$ is the hybridization amplitude.
Finally, the mean-field superconducting contribution to the Hamiltonian
is
\begin{eqnarray}
\mathcal{H}_{SC} & = & \sum_{k}\left\{ \Delta_{k}a_{k}^{\dagger}b_{-k}^{\dagger}+\Delta_{k}b_{k}^{\dagger}a_{-k}^{\dagger}\right.\nonumber \\
 &  & \left.+\Delta_{A,k}a_{k}^{\dagger}a_{-k}^{\dagger}+\Delta_{B,k}b_{k}^{\dagger}b_{-k}^{\dagger}+\text{h.c.}\right\} ,
\end{eqnarray}
with $\Delta_{k}=i\Delta\sin\left(k\right)$ where $\Delta$ is the
superconducting inter-band pairing amplitude, and $\Delta_{\text{(A,B)},k}=i\Delta_{\text{(A,B)}}\sin\left(k\right)$
where $\Delta_{\text{A}}$ and $\Delta_{\text{B}}$ are the superconducting
intra-band pairing amplitudes. 
We could also include a superconducting
term that changes Cooper pairs between different orbitals, which in terms of two particles interaction
may be written as $\sum_{k,k'}g_{J}\left(k,k'\right)\left(b_{k}^{\dagger}b_{-k}^{\dagger}a_{-k'}a_{k'}+a_{k}^{\dagger}a_{-k}^{\dagger}b_{-k'}b_{k'}\right)$,
where $g_{J}$ is the interaction strength. Without fluctuation, i.e.,
in the BCS theory, this term appears as an additive parameter to $\Delta_{\text{A}}$
and $\Delta_{\text{B}}$, thus besides enhancing the intra-band superconductivity
it does not change qualitatively the topological properties of the
Hamiltonian considered here.

In the more compact BdG form, the Hamiltonian may be written in the Nambu representation \cite{Nambu-1960} as $\mathcal{H}=\sum_{k}\boldsymbol{C}_{k}^{\dagger}\mathcal{H}_{k}\boldsymbol{C}_{k},$
where $\boldsymbol{C}_{k}^{\dagger}=\begin{pmatrix}a_{k}^{\dagger}b_{k}^{\dagger}a_{-k}b_{-k}\end{pmatrix}$
and
\begin{eqnarray}
\mathcal{H}_{k} & = & -\mu\Gamma_{z0}-\varepsilon_{k}\Gamma_{zz}+\Delta_{k}\Gamma_{yx}\nonumber \\
 &  & +\Delta_{\text{A},k}\frac{1}{2}\left(\Gamma_{y0}+\Gamma_{yz}\right)+\Delta_{\text{B},k}\frac{1}{2}\left(\Gamma_{y0}-\Gamma_{yz}\right)\nonumber \\
 &  & +V_{k}\cdot\mathbb{I},\label{eq:hamilt completo}
\end{eqnarray}
where $\Gamma_{ij}=\tau_{i}\varotimes s_{j},\quad\forall\;i,j=0,x,y,z$;
$\tau$ and $s$ are the Pauli matrices acting on particle-hole and
sub-band spaces, respectively, and $s_{0}=\tau_{0}$ are the $2\times2$
identity matrices. With respect to the Hamiltonian parameters: $V_{k}=-V_{\text{as},k}\boldsymbol{\Gamma}_{zy}$
if the hybridization is anti-symmetric or $V_{k}=V_{\text{s},k}\boldsymbol{\Gamma}_{zx}$
if the hybridization is symmetric.

In this section we present self-consistent results for the superconducting
parameters $\Delta$, $\Delta_{\text{A}}$ and $\Delta_{\text{B}}$
using the BdG formalism. The Hamiltonian defined in Eq. (\ref{eq:hamilt completo})
can be solved using BdG transformations as $a_{k}=\sum_{n}\left[u_{n,k}^{a}\gamma_{n,k}+\left(v_{n,k}^{a}\right)^{*}\gamma_{n,-k}^{\dagger}\right]$
and $b_{k}=\sum_{n}\left[u_{n,k}^{b}\gamma_{n,k}+\left(v_{n,k}^{b}\right)^{*}\gamma_{n,-k}^{\dagger}\right]$.
This transformation diagonalizes the Hamiltonian in the form $\mathcal{H}_{k}\psi_{n}=E_{n}\psi_{n}$,
with $\psi_{n}=\begin{pmatrix}u_{n,k}^{a} & u_{n,k}^{b} & v_{n,-k}^{a} & v_{n,-k}^{b}\end{pmatrix}_{n}^{T}$,
where $E_{n}$ are the energy eigenvalues and the wave function spinors
$\psi_{n}$ are the eigenstates.

The self-consistent solution implies that the pairings can be obtained
using
\begin{eqnarray}
\Delta & = & g\frac{1}{L}\sum_{k}i\sin\left(k\right)\left(\left\langle a_{k}b_{-k}\right\rangle +\left\langle b_{k}a_{-k}\right\rangle \right),\\
\Delta_{\text{A}} & = & g_{\text{A}}\frac{2}{L}\sum_{k}i\sin\left(k\right)\left\langle a_{k}a_{-k}\right\rangle ,\\
\Delta_{\text{B}} & = & g_{\text{B}}\frac{2}{L}\sum_{k}i\sin\left(k\right)\left\langle b_{k}b_{-k}\right\rangle ,
\end{eqnarray}
where $g$, $g_{\text{A}}$ and $g_{\text{B}}$ are the strength of the interactions
between fermions in different orbitals, in orbitals $A$ and in orbitals
$B$, respectively. At zero temperature, using the representation
of fermionic operators in terms of the Bogoliubov coefficients, we
may write
\begin{eqnarray}
\Delta & = & g\frac{1}{L}\sum_{k}\sum_{n}i\sin\left(k\right)\left[u_{n,k}^{a}\left(v_{n,-k}^{b}\right)^{*}+u_{n,k}^{b}\left(v_{n,-k}^{a}\right)^{*}\right],\nonumber \\
\\
\Delta_{\text{A}} & = & g_{\text{A}}\frac{2}{L}\sum_{k}\sum_{n}i\sin\left(k\right)u_{n,k}^{a}\left(v_{n,-k}^{a}\right)^{*},\\
\Delta_{\text{B}} & = & g_{\text{B}}\frac{2}{L}\sum_{k}\sum_{n}i\sin\left(k\right)u_{n,k}^{b}\left(v_{n,-k}^{b}\right)^{*}.
\end{eqnarray}

\subsection{Anti-symmetric hybridization}

We first consider the case of anti-symmetric hybridization ($V_{\text{as}}$)
that occurs when the orbitals angular momenta have different parities,
like orbitals $s$ and $p$. In Fig \ref{self-consist anti-sym hybrid}
we show the results for the three order parameters calculated self-consistently,
when $g/2=g_{\text{A}}=g_{\text{B}}=1.7$. 
A similar model was considered before \cite{Puel-Sacramento-Continentino-2015}
with only inter-band pairing.
The strength
of the coupling $g$ only changes the superconducting amplitude of
the SC phases (inter- or intra-band ones), thus its choice does not
change qualitatively the results presented. It is interesting to point
out that the self-consistent results for the superconducting order
parameters may converge to different results depending on the initial
guesses. This is a consequence of the first order nature of the quantum
phase transitions between the different ground states. Therefore it
is necessary to calculate the energy of the different states to obtain
the true ground state for a given set of parameters.

We note first that inter and intra-band superconductivity do not coexist
as equilibrium states. Their coexistence implies that one of them is
metastable. Second, we note that the intra-band SC does not distinguish
between different bands, in the sense that the results are equal for
both pairings. We note that considering any fixed value of the chemical
potential in the region where there is SC, when the anti-symmetric
hybridization is increased it eventually destroys the inter-band SC
that is present. On the other hand, if we keep increasing the hybridization,
it raises the intra-band SC up to a maximum value until it suppresses
the SC definitely.

In Fig. \ref{self-consist anti-sym hybrid}d we show the spectral gap for
the self-consistent results.
Also we show the phase diagram in the right plot of the same figure.
As we can see, the consideration
of inter-band, intra-band superconductivity and anti-symmetric hybridization
results in a rich phase diagram. In this figure, the solid lines represent
a gap closing, while the dashed lines represent a phase separation
without closing the gap. Phase I in this figure is a gapless superconducting
phase, driven by the inter-band coupling, and it was shown \cite{Puel-Sacramento-Continentino-2015}
to behave like Weyl superconductor. The phase II is a two-band superconductor
with only intra-band couplings. Phase III is a topological insulator
which was shown to have localized states at the edges \cite{Puel-Sacramento-Continentino-2015}
of a finite chain. The phase IVa shows gapped superconductivity and
represents the strong inter-band coupling superconducting phase. The
phase IVb is a trivial insulator and there is no SC remaining. Finally,
phase V is a normal metallic phase. All those phases are symmetric
around $\mu=0$.
Since the intra- and inter-band pairings do not coexist, the phases with
no intra-band pairing are similar to the results previously obtained \cite{Puel-Sacramento-Continentino-2015}. The main difference
results from the appearance of the intra-band pairing in some regions of the
phase diagram.

\subsection{Symmetric hybridization}

Analogously to the previous case, we also calculate the order parameters
self-consistently considering symmetric hybridization ($V_{\text{s}}$).
This is the case when the orbitals angular momenta have equal parities,
like orbitals $s$ and $d$. In Fig. \ref{self-consist sym hybrid}
we show the results for the same set of values $g$, $g_{\text{A}}$
and $g_{\text{B}}$ as the anti-symmetric case. First, we notice that
the intra-band SC distinguishes between different bands,
since there is a change of sign between them. 
Unlike the anti-symmetric
case, here there is a coexistence of inter- and intra-band SC. Remarkably,
the inter-band has the larger order parameter for all region of parameters.
In general, this indicates that the inter-band SC has higher critical
temperature, which turns out to be responsible for the superconductivity
appearing in the material. Note that symmetric hybridization is responsible
for the emergence of intra-band SC. Very strong symmetric hybridization
eventually destroys superconductivity.

In Fig. \ref{self-consist sym hybrid} we also show the spectral
gap for the self-consistent
results. 
We also show the
phase diagram in the right plot of Fig. \ref{self-consist sym hybrid},
as in Fig. \ref{self-consist anti-sym hybrid}.
As before, the solid
lines represent a gap closing, while the dashed lines represent a
phase separation without closing the gap. Phase I and IIa are gapped
superconducting phases, with the coexistence of inter- and intra-band
couplings, but dominated by the inter-band one. Phase IIb is an insulating
phase and there is no SC. All those phases are symmetric around $\mu=0$.
The more interesting phase is phase I, which allows both types of
couplings and shows non-trivial topological properties. This phase
is characterized by localized edge states and finite winding number, as will be shown in the
next section.

The robustness of the inter-band superconductivity can be tested
varying the relative amplitudes of the $g$, $g_{A}$ and $g_{B}$
parameters. Considering, for instance, the case $g_{A}=g_{B}\equiv g_{0}$
and selecting the point $\mu=0$ and $V_{s}=1$,
the appearance of the inter-band SC is not continuous
with increasing $g$, but goes through a first order transition at
some point $g>g_{0}$ near to $g = g_0$ to a value that
always has a larger amplitude than the intra-band ones. 
While the results of Figs. \ref{self-consist sym hybrid} consider a large $g$ value,
the results are qualitatively the same, as long as the inter-band SC
is present.

\section{Topological properties \label{sec:Topology}}

\subsection{Winding number in the BDI class}

The symmetry-protected topological systems are classified accordingly
to their symmetries \cite{Schnyder-Ryu-Furusaki-Ludwig-2008}. The Hamiltonian
of equation (\ref{eq:hamilt completo}) has particle-hole symmetry once
it obeys the relation $\mathcal{H}_{k}=-\mathcal{O}\mathcal{H}_{k}\mathcal{O}^{-1}$ ~ \cite{Schnyder-Ryu-Furusaki-Ludwig-2008},
where the operator written in the Nambu representation \cite{Nambu-1960} is $\mathcal{O}=\boldsymbol{\Gamma}_{x0}K$,
in which $\mathcal{O}^{2}=+1$ and $K$ applies the complex conjugate
and inverts the momentum. In addition, the Hamiltonian has simplified
time reversal symmetry for spinless fermions, $\mathcal{H}_{k}=\mathcal{H}_{-k}^{*}$.
In the presence of both symmetries, the Hamiltonian belongs to the
BDI class of topological systems, and the one-dimensionality guarantees
that the space of the quantum ground state is partitioned into topological
sectors labeled by an integer ($\mathbb{Z}$) number \cite{Schnyder-Ryu-Furusaki-Ludwig-2008}.

In the $\mathbb{Z}$ class of topological systems,
the topological phases in odd-dimensional systems (or, in other words, those with chiral symmetry) are characterized
by the topological invariant called winding number \cite{Schnyder-Ryu-Furusaki-Ludwig-2008,Yada-Sato-Tanaka-Yokoyama-2011}.
This invariant counts the number of the zero-energy states protected
by the topological property of the Hamiltonian, and may be calculated
in the usual way \cite{Yada-Sato-Tanaka-Yokoyama-2011,Ii-Yada-Sato-Tanaka-2011}.
One needs to look for an hermitian matrix which anti-commutes
with the Hamiltonian $\left(\mathcal{H}\right)$, i.e., find $\Gamma$
such that $\left\{ \mathcal{H},\Gamma\right\} =0.$
Considering spinless time-reversal symmetry and
particle-hole symmetry (PHS)
then the Chiral operator that carries both symmetries is $\Gamma_{x0}$.
It implies that the Hamiltonian
anti-commutes with that operator, which can be used to bring the Hamiltonian to an off-diagonal form.
Using the basis that diagonalizes $\Gamma_{x0}$, i.e., $R^{-1}\;\Gamma_{x0}\;R=D$, with
$R=\Gamma_{xx}-\Gamma_{zx}$ and $D$ a diagonal matrix,
implies that
\begin{equation}
R^{-1}\;\mathcal{H}_{k}\;R=\begin{pmatrix}0 & q\left(k\right)\\
q^{\dagger}\left(k\right) & 0
\end{pmatrix}.
\end{equation}
Writing a generic Hamiltonian in the form
\begin{eqnarray}
\mathcal{H}_{k} & = & \sum_{i,j}h_{ij}\boldsymbol{\Gamma}_{ij},\qquad i,j=0,x,y,z,\label{eq:HGamma}
\end{eqnarray}
whose coefficients $h_{ij}$ may be extracted from any generic Hamiltonian
${\cal H}$ through $h_{ij}=\frac{1}{4}\text{Tr}\left(\boldsymbol{\Gamma}_{ij}\mathcal{H}\right)$,
if we apply the PHS to Eq. (\ref{eq:HGamma}) as $\mathcal{H}_{k}=-\Gamma_{x0}\mathcal{H}_{-k}^{T}\Gamma_{x0}$
and proceed with the block off-diagonal calculations described above
we find that
\begin{eqnarray}
q\left(k\right) & = & \sum_{j}c_{j}\left(h_{zj}+ih_{yj}\right)\sigma_{j},\qquad j=0,x,y,z,
\end{eqnarray}
where $c_{0}=c_{x}=+1$ and $c_{y}=c_{z}=-1$, $\sigma_{x,y,z}$ are
the Pauli matrices and $\sigma_{0}$ is the $2\times2$ identity matrix.

The winding number, $W$, is defined as the number of revolutions
of $\det\left[q\left(k\right)\right]=m_{1}\left(k\right)+im_{2}\left(k\right)$
around the origin in the complex plane when $k$ changes from $-\pi$
to $\pi$,
\begin{equation}
W=\frac{1}{2\pi}\int_{-\pi}^{\pi}\frac{\partial\theta\left(k\right)}{\partial k}dk,\label{eq:winding number}
\end{equation}
with 
\begin{equation}
\theta\left(k\right)=\arg\det\left[q\left(k\right)\right]=\tan^{-1}\frac{m_{2}\left(k\right)}{m_{1}\left(k\right)}.
\end{equation}
For the generic case considered above we have that
\begin{eqnarray}
m_{1}\left(k\right) & = & \sum_{j}d_{j}\left(h_{zj}^{2}-h_{yj}^{2}\right)\nonumber \\
 & \text{and}\nonumber \\
m_{2}\left(k\right) & = & \sum_{j}d_{j}\left(2h_{zj}h_{yj}\right),\label{m1 and m2}
\end{eqnarray}
where $d_{0}=+1$ and $d_{x,y,z}=-1$.

\subsection{Edge states in a finite chain \label{sec:Energy spectrum}}

In order to find the energy spectrum of a finite chain of fermions
through the BdG transformation we write the Hamiltonian, Eq. (\ref{eq:hamilt completo})
transformed to real space, in the form
\begin{equation}
\mathcal{H}=\bm{C}^{\dagger}\boldsymbol{H}\bm{C},
\end{equation}
where
\begin{equation}
\boldsymbol{C}=\begin{pmatrix}a_{1} & b_{1} & a_{1}^{\dagger} & b_{1}^{\dagger} & \cdots & a_{N} & b_{N} & a_{N}^{\dagger} & b_{N}^{\dagger}\end{pmatrix}^{T}
\end{equation}
and the operators $a_{i}^{\dagger}\left(a_{i}\right)$ and $b_{i}^{\dagger}\left(b_{i}\right)$
create (annihilate) a fermion in the orbital A and B, respectively,
at position $i$ in the chain. The matrix $\boldsymbol{H}$ is defined
as
\begin{equation}
\boldsymbol{H}=\begin{pmatrix}\mathcal{H}_{11} & \cdots & \mathcal{H}_{1N}\\
\vdots & \ddots & \vdots\\
\mathcal{H}_{N1} & \cdots & \mathcal{H}_{NN}
\end{pmatrix},
\end{equation}
and is comprised by the following $\left(4\times4\right)$ interaction
matrices
\begin{equation} \label{Hamilt_terms_real_space}
\begin{cases}
\mathcal{H}_{r,r} & =-\mu\Gamma_{z0},\\
\mathcal{H}_{r,r+1} & =-t\Gamma_{zz}-i\frac{\Delta}{2}\Gamma_{yx}-i\frac{\Delta_{0}}{2}\Gamma_{y0}+V\left(r+1\right),\\
\mathcal{H}_{r,r-1} & =-t\Gamma_{zz}+i\frac{\Delta}{2}\Gamma_{yx}+i\frac{\Delta_{0}}{2}\Gamma_{y0}+V\left(r-1\right),\\
\mathcal{H}_{r,r'} & =0\qquad\forall\;r'\neq r,\;r+1\;\text{or}\;r-1,
\end{cases}
\end{equation}
where $V\left(r+1\right)=-V\left(r-1\right)=-i\frac{V}{2}\Gamma_{zy}$
for anti-symmetric hybridization, and $V\left(r+1\right)=V\left(r-1\right)=\frac{V}{2}\Gamma_{zx}$
for symmetric one.

If we consider the BdG transformation as the following
\begin{equation}
\begin{array}{c}
a_{r}=\sum_{n}\left[u_{s,n}(r)\gamma_{n}+v_{s,n}^{*}(r)\gamma_{n}^{\dagger}\right],\\
b_{r}=\sum_{n}\left[u_{p,n}(r)\gamma_{n}+v_{p,n}^{*}(r)\gamma_{n}^{\dagger}\right],
\end{array}
\end{equation}
it diagonalizes the Hamiltonian, $\mathcal{H}=E_{0}+\sum_{n}E_{n}\gamma_{n}^{\dagger}\gamma_{n}$,
such that $\boldsymbol{U}^{\dagger}\boldsymbol{H}\boldsymbol{U}=\boldsymbol{E}$,
where $\boldsymbol{U}$ is formed by all the BdG coefficients $u_{s}$,
$v_{s}$, $u_{p}$ and $v_{p}$, and has the property to be unitary
$\boldsymbol{U}^{\dagger}\boldsymbol{U}=\mathbb{I}$. The matrix $\boldsymbol{E}$
is diagonal and contains the energy spectrum ($E_{n}$) of the system.

\subsection{$4\pi$ Josephson effect \label{sec:Josephson}}

\begin{figure}
\includegraphics[width=0.4\textwidth]{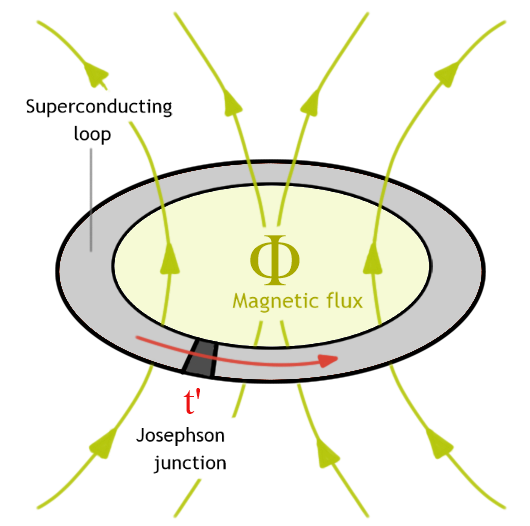}
\caption{Schematic figure illustrating the 1D superconducting ring with a Josephson junction.}
\label{fig: josephson junction}
\end{figure}

In the previous section we have considered a 1D open chain, i.e., there is
no connection between sites $1$ and $N$. In terms of
eq. (\ref{Hamilt_terms_real_space}) we have $\mathcal{H}_{N,1}=\mathcal{H}_{1,N}=0$.
Now we may think of a chain as a ring with a Josephson junction coupling the ends, see Fig. \ref{fig: josephson junction}.
An extra hopping term $t^{\prime}$ couples the end point of the ring to the first point
via some insulating junction.
If a uniform magnetic field ($\Phi$) flows through this ring,
its effect may be captured by a Peierls substitution in the extra hopping term, $t^{\prime}$
\cite{Garcia-LeBlanc-Williams-Beeler-Perry-Spielman-2012}. Thus, the Josephson junction may be represented by the following boundary conditions
\begin{equation}
\mathcal{H}_{N,1}	= \mathcal{H}_{1,N}^{*} =
\begin{pmatrix}-\text{e}^{-i\phi/2}t' & 0 & 0 & 0\\
0 & \text{e}^{-i\phi/2}t' & 0 & 0\\
0 & 0 & \text{e}^{i\phi/2}t' & 0\\
0 & 0 & 0 & -\text{e}^{i\phi/2}t'
\end{pmatrix},
\end{equation}
where the superconducting phase difference $\phi$
across the junction is related to the magnetic flux through the ring
by $\phi=2\pi\Phi/\Phi_{0}$, and $\Phi_{0}=h/2e$
is the superconducting flux quantum.
We have that $t^{\prime}$ is the tunneling,
or inversely proportional to a barrier amplitude,
across the junction. As mentioned above this is equivalent to the original
proposal of the Josephson junction between two different superconductors with different
pairing phases also separated by some tunneling amplitude accross an insulator (or metal).

We may now analyze the junction effect on a current flowing in the ring
as we change the magnetic flux by discrete amounts of flux quantum,
by changing the junction phase $\phi$ by multiples of $2\pi$.
In a normal superconductor each additional flux quantum ($\Delta \phi = 2\pi$, usually called a pump) 
should lead the system to its initial state \cite{Byers-Yang-1961}.
On the other hand, the topological superconductor (TSC) changes its parity at every pump \cite{Pientka-Romito-Duckheim-Oreg-Oppen-2013},
leading the system to a different final state after pumping.
The reason is that the TSC is allowed to have zero energy crossings in its spectrum of excitation during the pump
and therefore only returns to its initial state after a further change of the phase by $2\pi$.

\begin{figure}
\includegraphics[width=0.5\textwidth]{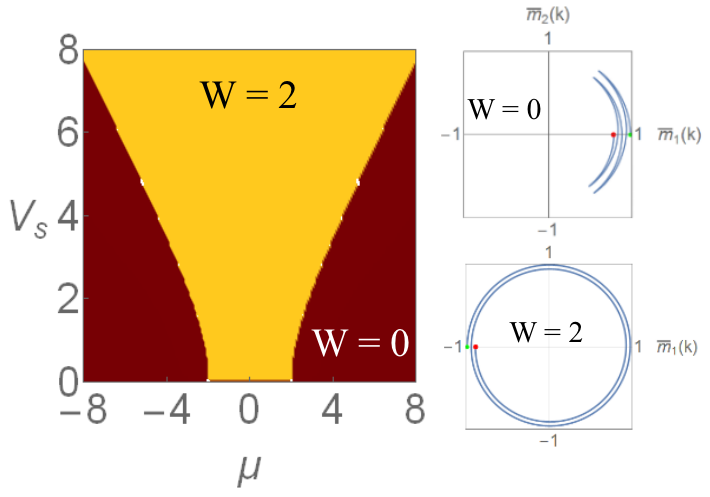}
\caption{
In the left panel we show the winding number calculated from the self-consistent results
for symmetric hybridization, over the phase space of parameters.
In the right panels we show the normalized parametric plot of real and imaginary
parts of $\det\left[q\left(k\right)\right]$. 
The number of times $\det\left[q\left(k\right)\right]$
wraps the origin is the winding number and is illustrated in the right side.
}
\label{fig:Winding Number}
\end{figure}

\begin{figure}[th]
\includegraphics[width=0.35\textwidth]{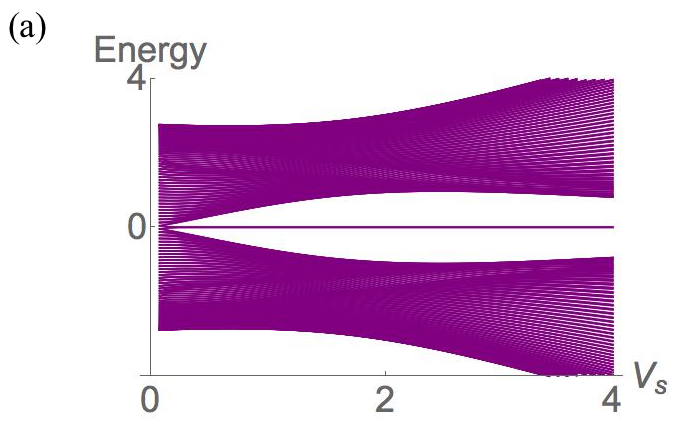}
\includegraphics[width=0.35\textwidth]{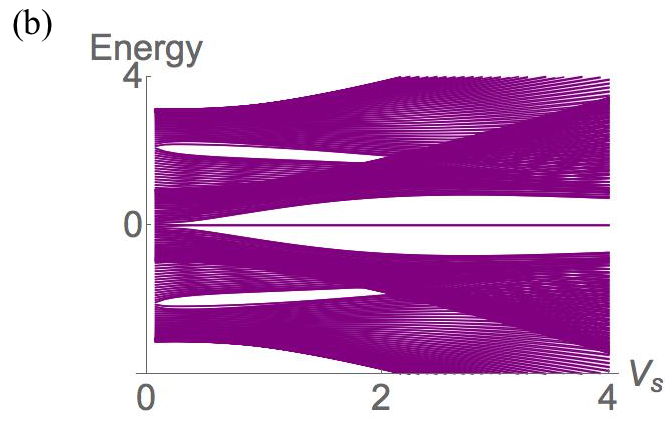}
\caption{Here we show the energy spectrum of the self-consistent
results, for two fixed values of the chemical potential and increasing
symmetric hybridization ($\mu=0$ on (a) and $\mu=-1.4$
on (b)).}
\label{fig:energy spectrum symmetric}
\end{figure}

\begin{figure*}[th]
\includegraphics[width=0.32\textwidth]{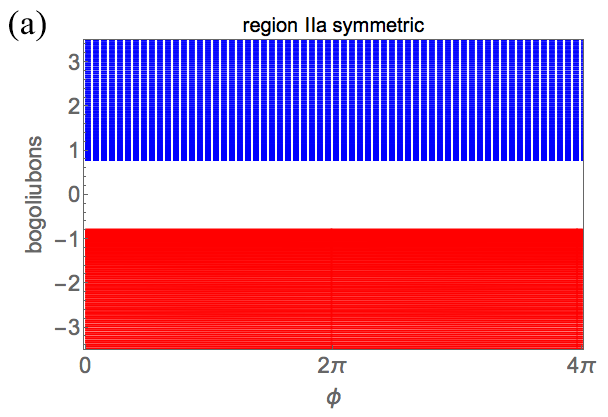}
\includegraphics[width=0.32\textwidth]{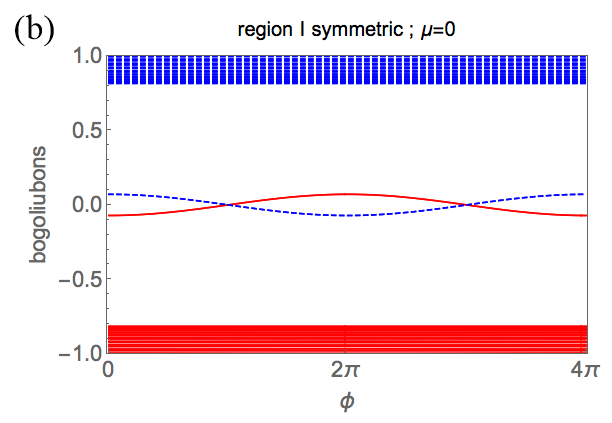}
\includegraphics[width=0.32\textwidth]{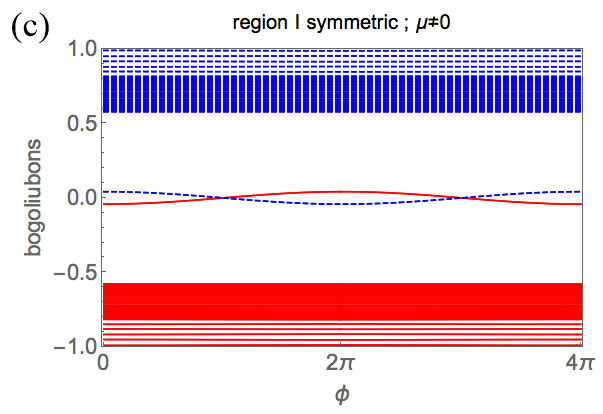}

\includegraphics[width=0.32\textwidth]{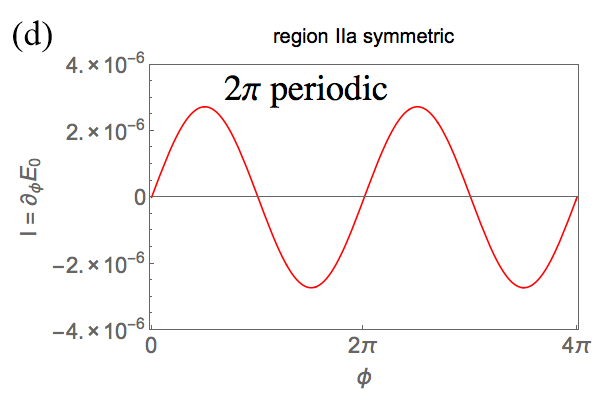}
\includegraphics[width=0.32\textwidth]{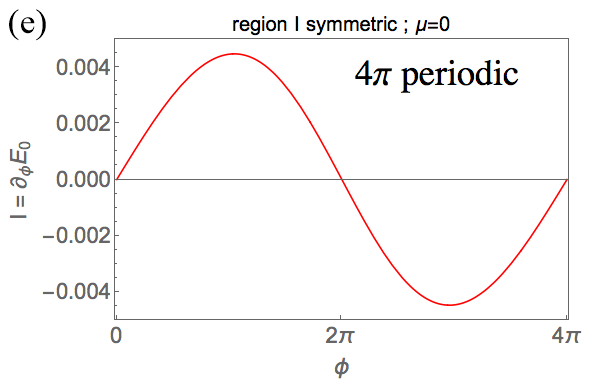}
\includegraphics[width=0.32\textwidth]{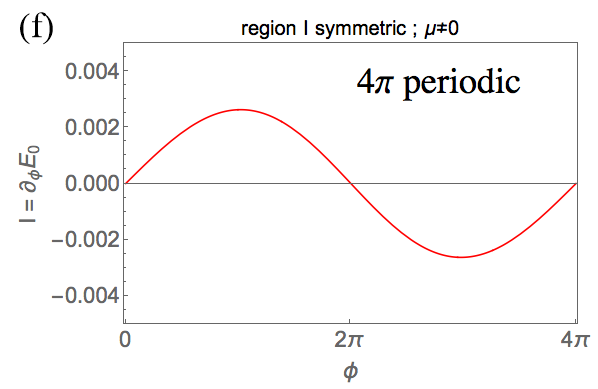}
\caption{Results for the case of symmetric hybridization case as we vary the tunneling phase $\phi$: i) First row 
shows the excitation spectrum that preserves the parity of the superconductor. ii) Second row shows the Josephson 
current through the Josephson junction. Here we have used $L = 150$ and $t^{\prime} = 0.1$.}
\label{fig: pi Josephson current symmetric}
\end{figure*}

\subsection{Symmetric hybridization}

\paragraph{Winding number:}
we begin our analysis of the topological phases of the proposed model with the winding number calculation.
For convenience,
we'll consider the case where $\Delta_{\text{B}}=-\Delta_{\text{A}}=\Delta_{0}$.
If we compare Eq. (\ref{eq:hamilt completo}) -- with symmetric hybridization $V_{s,k}$ --
and Eqs. (\ref{m1 and m2}) we have 
$m_{1}\left(k\right)=\mu^{2}+\Delta_{k}^{2}+\Delta_{0,k}^{2}-V_{\text{s},k}^{2}-\epsilon_{k}^{2}$
and $m_{2}\left(k\right)=-2\left(V_{\text{s},k}\Delta_{k} - \epsilon_{k}\Delta_{0,k} \right)$.
This suggests that the symmetric hybridization may induce a topological
phase, since we have non-vanishing $m_{2}$ even to zero chemical potential.
To be sure that the phase is topological we must calculate
the winding number itself, or see if the parametric plot of $\bar{m}_{1}\left(k\right)$
and $\bar{m}_{2}\left(k\right)$ contains the origin when $k\in\left[-\pi,\pi\right]$.
The results for the winding number and the parametric plot are 
shown in Fig. \ref{fig:Winding Number}
for the parameters $V_s=1.2$, $\mu=-1.04$.  
This figure shows that the parametric plot wraps the origin twice; it
means that the winding number in this case is two, $W=2$. 
The results for the winding number clearly show the topological
phase, induced by symmetric hybridization, and dominated by inter-band
superconductivity for small values of the chemical potential
that grows as the hybridization, $V_s$, grows.

\paragraph{Edge states --}
Since we have defined the topological region of the parameters,
we may analyse the zero-energy modes explicitly through the energy spectrum of a finite chain.
We have calculated the energy spectrum for a chain of $L=100$ sites,
therefore, we get $4L$ energies for the spectrum. We have checked
that this size is large enough to prevent finite size effects. We
analyze the energy spectrum for two fixed values of chemical potential,
$\mu=0$ and $\mu=-1.4$, and increasing the hybridization according
to the self-consistent solution of Fig. \ref{self-consist sym hybrid}.
The results are shown in Fig. \ref{fig:energy spectrum symmetric}.
What we immediately see is that the zero-energy states are robust,
i.e., even when $\mu$ is non-zero they are present,
which characterizes the zero-energy modes in the superconducting phase.
We notice that those states are four-fold degenerated. 
We have checked that they have wavefunctions
that are localized exponentially close to the edges if the system is large enough.

\paragraph{$4\pi$ Josephson effect:}
we may also analyse the topological properties of the system via Josephson junction scheme, see Fig. \ref{fig: josephson junction}.
First, we look to the excitation spectrum (bogoliubons) during two pumps
for each superconducting phase in the phase diagram.
The results are shown in the first row of Fig. \ref{fig: pi Josephson current symmetric},
where \ref{fig: pi Josephson current symmetric}a is for the trivial phase IIa,
whereas \ref{fig: pi Josephson current symmetric}b and \ref{fig: pi Josephson current symmetric}c are for the topological phase I
for two values of the chemical potential.
We may see that there are level crossings when the SC is in its topological phase
and there is no crossing in the trivial one.

To explicitly see the periodicity of the Josephson current during the pump, 
we need to analyse the ground state energy ($E_0$) of the superconductor preserving its parity, i.e.,
the ground state is composed by the solid (red) lines of the excitation spectrum. Dashed (blue) lines carry the opposite parity. 
Thus, the sum over the "negative" excitation to compute $E_0$ needs to follow the excitation when it crosses the zero energy state.
In the topological phase, the crossing through zero energy is a direct consequence of the presence of the zero energy mode at 
the end of the chain. Here we have two zero energy excitations at each end, thus it is natural that we have two 
level crossings (we notice that region I with $\mu = 0 $ in Fig. \ref{fig: pi Josephson current symmetric} has a degenerate level crossing).
When $\mu \neq 0$ the level crossing modes do not need to be degenerated,
but we notice that even though we have two level crossings (and their particle-hole symemtric), the crossings through zero always happen at the same $\phi$ point.
Second row of Fig. \ref{fig: pi Josephson current symmetric} shows the current flowing through the junction, which is the derivative of the ground state energy respective to the flux $\phi$.
We clearly see that the current has a periodicity of $2\pi$ (one pump) in the trivial phase, 
Fig. \ref{fig: pi Josephson current symmetric}d.
On the other hand, the periodicity of the Josephson currents in
Figs. \ref{fig: pi Josephson current symmetric}e and \ref{fig: pi Josephson current symmetric}f
are $4\pi$ (two pumps), characterizing the topological superconducting phase
and providing an alternative evidence for the presence of Majorana states.

\begin{figure}[th]
\includegraphics[width=0.35\textwidth]{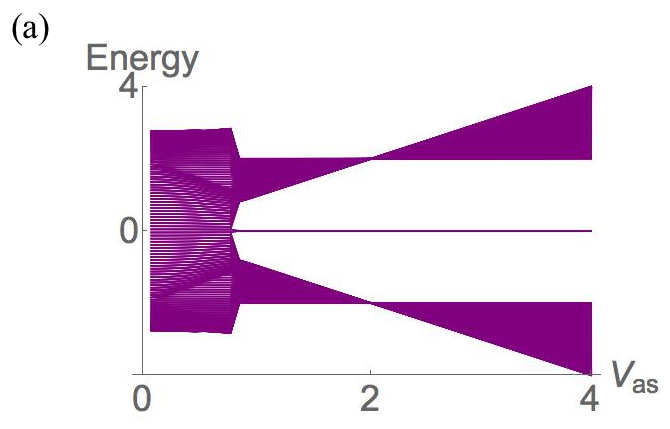}
\includegraphics[width=0.35\textwidth]{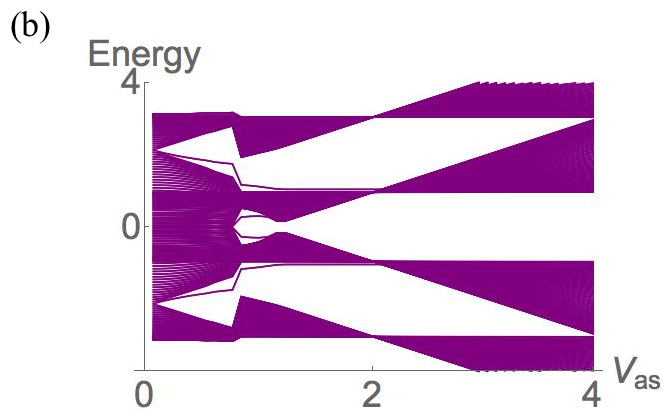}
\caption{Here we show the energy spectrum of the self-consistent
results, for two fixed values of the chemical potential and increasing
anti-symmetric hybridization ($\mu=0$ on (a) and $\mu=-1.4$
on (b)).}
\label{fig:energy spectrum}
\end{figure}

\begin{figure*}[th]
\includegraphics[width=0.32\textwidth]{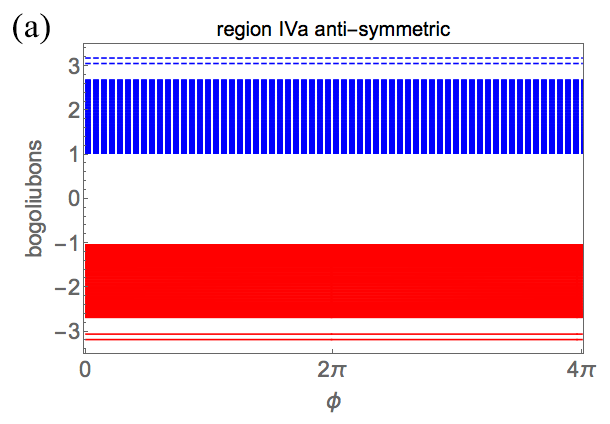}
\includegraphics[width=0.32\textwidth]{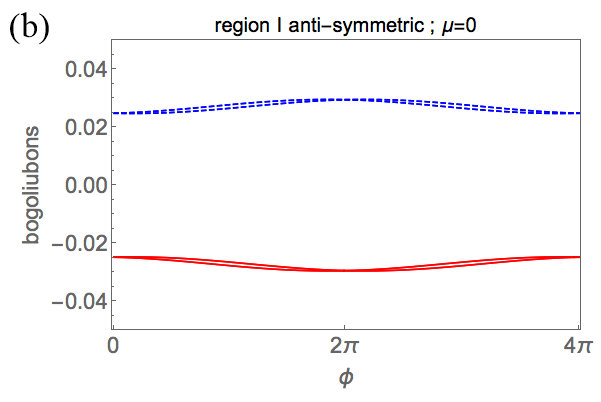}
\includegraphics[width=0.32\textwidth]{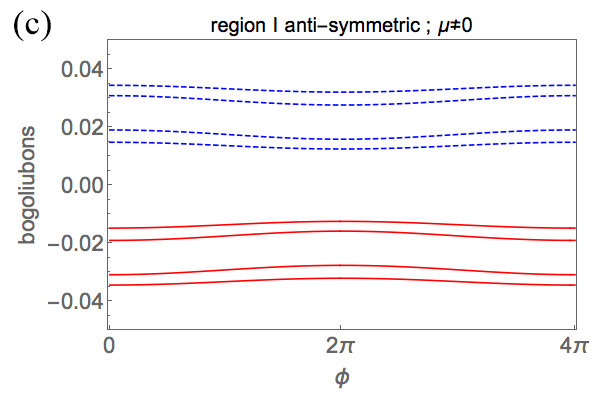}

\includegraphics[width=0.32\textwidth]{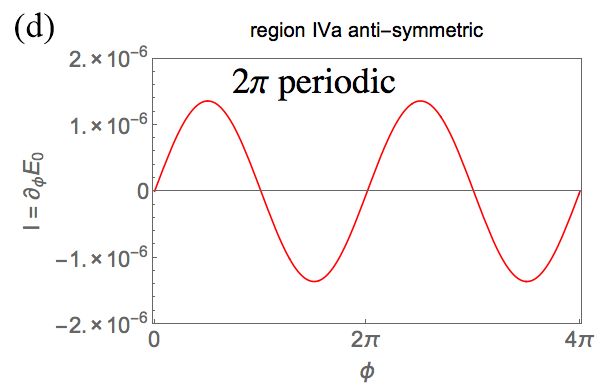}
\includegraphics[width=0.32\textwidth]{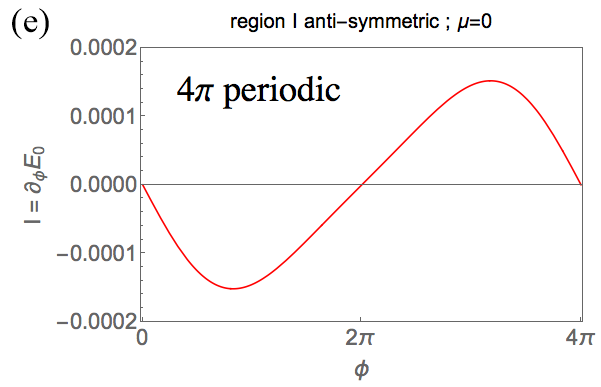}
\includegraphics[width=0.32\textwidth]{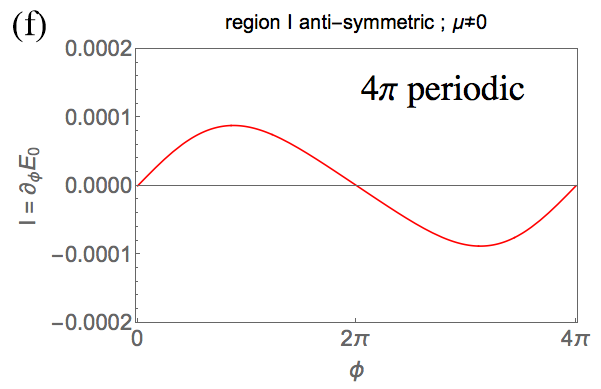}
\caption{Results for anti-symmetric hybridization as we vary the tunneling phase $\phi$: i) First row shows the 
excitation spectra that preserve the parity of the superconductor. ii) Second row shows the Josephson current 
flowing through the Josephson junction. Here we have used $L = 250$ and $t^{\prime} = 0.1$.}
\label{fig: pi Josephson current anti-symmetric}
\end{figure*}

\subsection{Anti-symmetric hybridization}

\paragraph{Winding number:}
we proceed the analysis of the topological properties with the winding number calculation.
For convenience, and since
the self-consistent results do not distinguish the SC in the bands,
we'll consider the case where $\Delta_{\text{A}}=\Delta_{\text{B}}=\Delta_{0}$.
Therefore, comparing Eq. (\ref{eq:hamilt completo}) -- with anti-symmetric hybridization, $V_{as}$ -- and Eqs. (\ref{m1 and m2})
we have that $m_{1}\left(k\right)=\mu^{2}+\Delta_{k}^{2}-\Delta_{0,k}^{2}-V_{\text{as},k}^{2}-\epsilon_{k}^{2}$
and $m_{2}\left(k\right)=-2\mu\Delta_{0,k}$. As a result we notice
that only for a non-zero chemical potential and intra-band superconductivity
we have non-vanishing $m_{2}$ and the system may include a topological
phase. 
Calculating
the winding number, as described in Eq. (\ref{eq:winding number}),
one obtains a trivial solution ($W=0$) for all self-consistent
solutions in parameter space \cite{Puel-Sacramento-Continentino-2015}.
Even though the winding number seems to indicate a trivial solution, 
the results of Ref. [\onlinecite{Puel-Sacramento-Continentino-2015}]
for a system with no intra-band pairing
show that the phases corresponding to regions I and III of the phase
diagram in Fig. \ref{self-consist anti-sym hybrid} are topological. The
topological property of phase III is hidden by particle-hole symmetry.
Moreover, Ref. [\onlinecite{Puel-Sacramento-Continentino-2015}]
shows that in this phase localized states are present in the edges
of the chain (despite having finite energy when $\mu\neq0$). As concerns
phase I, it is a topological phase that presents Weyl fermions \cite{Puel-Sacramento-Continentino-2015}
whose topological character remains also undetected by the winding
number calculation. In that reference it is shown an alternative procedure
to uncover the topological nature of this phase.

\paragraph{Edge states --}
now we proceed with the analysis of the zero-energy modes explicitly through the energy spectrum of a finite chain.
We also have used $L=100$ sites,
which is large enough to prevent finite size effects. We
analyze the energy spectrum for two fixed values of chemical potential,
$\mu=0$ and $\mu=-1.4$, and increasing the hybridization according
to the self-consistent solution of Fig. \ref{self-consist anti-sym hybrid}.
The results for anti-symmetric
hybridization are shown in Fig. \ref{fig:energy spectrum}.
We immediately see that the zero-energy states for $\mu=0$ are not
robust, in the sense that they disappear when $\mu\neq0$.
This is the difference of zero-energy modes in the superconductor (phase I for symmetric hybridization)
and zero-energy modes in the insulator (phase III for the anti-symmetric hybridization).
The chemical potential is not breaking any symmetry,
but the zero-energy modes in the superconductor are topologically protected and survive after the introduction of a finite $\mu$,
while in the insulator those zero-energy modes are not protected and can be eliminated as you see in this figure.

\paragraph{$4\pi$ Josephson effect:}

we may also analyse the topological properties of the system via Josephson junction scheme (Fig. \ref{fig: josephson junction}).
We start looking to the excitation spectrum (bogoliubons) during two pumps
for each superconducting phase in the phase diagram.
The anti-symmetric case has three types of superconducting phases: intraband gapped SC, interband gapped SC 
and interband gapless SC, as shown in Fig. \ref{self-consist anti-sym hybrid}e.
Both gapped superconducting phases (II and IVa) show similar excitation spectra and their typical bogoliubons 
that keeps the ground state parity are shown in Fig. \ref{fig: pi Josephson current anti-symmetric}a.
As expected, there are no level crossings in the excitation spectrum and
the current is $2\pi$ periodic as we can see in Fig. 
\ref{fig: pi Josephson current anti-symmetric}a for the case of region IVa.

In phase I, even though we have no gap in the bulk spectrum of an infinite system, it is still possible to calculate the Josephson current in a finite one.
The junction itself opens up a small gap in the spectrum if $L$ is not too large and $t^{\prime}$ is not too strong.
Of course, in the limit $L \rightarrow \infty$ the gap closes, but if the tunneling $t^{\prime}$ is too large 
(or the barrier too small) the junction just couples both ends analogously to a periodic boundary condition (i.e., infinite system).
Thus, a typical excitation spectrum for very small energies in the gap generated by the coupling
accross the junction (positive and negative excitation) is shown in 
Figs. \ref{fig: pi Josephson current anti-symmetric}b and \ref{fig: pi Josephson current anti-symmetric}c.

Even though Figs. \ref{fig: pi Josephson current anti-symmetric}b and \ref{fig: pi Josephson current anti-symmetric}c show no level crossings during the pumps, 
we may proceed with the same calculations as before and obtain the Josephson current.
The result is shown in Figs. \ref{fig: pi Josephson current anti-symmetric}e and \ref{fig: pi Josephson current anti-symmetric}f for two values of the chemical potential.
Clearly, both figures exhibit $4\pi$ periodic Josephson current,
even without zero energy level crossings revealing in some sense the hidden topological nature of
this Weyl-phase.


\section{Conclusions\label{sec:Conclusions}}

In this paper we have studied a model of a $p$-wave, one dimensional,
multiband superconductor. This represents a generalization of the
single band model for odd-parity superconductivity that gives rise
to a much richer phase diagram with a variety of quantum phase transitions.
The odd-parity superconductivity is preserved in this extension, but
inter-band superconductivity is now present in addition to the intra-band
ones. 
The presence of two-bands in our model allows
us to include hybridization, increasing the space of parameters. We
have considered symmetric and anti-symmetric hybridizations. Both
are permitted, depending on the parities we choose for the angular
momenta of the two orbitals.

We have calculated the self-consistent solutions for the inter- and
intra-band superconducting order parameters as functions of the chemical
potential and the strength of the symmetric or anti-symmetric hybridization.
The self-consistent calculation of the order parameters allow to obtain the $T=0$ phase diagram of the system.
When increasing anti-symmetric hybridization, both intra- and inter-band
superconductivity emerge in the phase diagram, but they compete and
exclude one another for different values of band-filling. On the other
hand, when increasing the symmetric hybridization, both types of superconductivity
are present and they coexist. An interesting result is that inter-band
superconductivity has the highest value of order parameter, indicating
that it has the higher critical temperature and makes it responsible
for the superconductivity appearing in the system.

A general approach for obtaining the winding number of
a system described by $4\times4$ matrices was presented. It may be
applied whenever particle-hole symmetry and spinless time-reversal symmetry are present in a
Bogoliubov-de Gennes (BdG) Hamiltonian, which is the case
of the two-bands BCS superconductors studied here. According
to this approach, a dominant inter-band coupling with symmetric hybridization
between bands induces a topological superconducting phase. The non-trivial
topological character of this phase was shown through a calculation
of the winding number, using the self-consistent solutions for the
different order parameters. In order to further clarify our results
concerning the nature of the topological phases and their end states,
we have analyzed the energy spectrum of a finite system. We have compared
the energy spectrum between the anti-symmetric and symmetric results,
or the trivial and topological results, respectively. We also checked
the localization of the zero-energy states.

In order to provide further evidence for the presence of edge
Majorana states we have shown that in the topological phases one finds a $4\pi$-periodic (fractional)
Josephson current as one changes the magnetic flux accross a ring composed of the
superconductor with an insulator inserted between its ends. The result is consistent
with the results for the winding number and edge states for the topological phase
in the case of symmetric hybridization. In addition, we also found the same
$4\pi$-periodic Josephson current in the hidden topological phase identified previously
as Weyl-type in the case of anti-symmetric hybridization.

As a final note, we highlight that symmetric hybridization in addition
to odd-parity inter-band superconductivity stabilizes a topological
non-trivial phase, which presents localized states at the ends of
the chain.

\section*{Acknowledgments}
The authors would like to thank the CNPq and FAPERJ for financial
support. They also are grateful to Emilio Cobanera for discussion and calling attention to Ref. [\onlinecite{Alase-Cobanera-Ortiz-Viola-2016}], and Griffith M.A.S. for useful discussions. Partial support from FCT through grant UID/CTM/04540/2013 is acknowledged.

\section*{References}

\bibliographystyle{apsrev4-1}
\bibliography{refs}

\end{document}